\documentclass[useAMS,usenatbib]{aa}  

\usepackage{graphicx}
\usepackage{txfonts}
\usepackage{hyperref}
\usepackage{longtable}
\usepackage{appendix}
\begin{document}

   \title{In space there will be no need to scream}

   \subtitle{Limits to the presence of giant planets in the $\zeta^2$ Ret system }

   \author{A.~Su{\'a}rez~Mascare{\~n}o\inst{\ref{iac},\ref{uiac}}}

    \authorrunning{A.~Su{\'a}rez Mascare{\~n}o et al.}
    
   \institute{Instituto de Astrof\'{\i}sica de Canarias, c/ V\'ia L\'actea s/n, 38205 La Laguna, Tenerife, Spain\label{iac}\\
    \email{asm@iac.es} 
    \and Departamento de Astrof\'{\i}sica, Universidad de La Laguna, 38206 La Laguna, Tenerife, Spain \label{uiac}
    }

   \date{Written October 2025}

% \abstract{}{}{}{}{} 
% 5 {} token are mandatory
 
  \abstract
   {The search for life beyond our Solar system has been a long and difficult endeavour. The majority of current efforts are focused on the potential detection of biosignatures. However, their detection and interpretation are extremely challenging. Technosignatures appear as an attractive alternative, given their expected univocal interpretation. In recent years, the number of publications discussing them have skyrocketted, both in their more rigurous and speculative sides. In this article, we explore the 28.8 years of archival radial velocity data of $\zeta^2$ Ret with the aim of detecting the proposed giant planet Calpamos, suspected source of a signal of technological origin. We performed a global model fitting the radial velocity data along with activity indicators and modelled the stellar magnetic cycle and rotation. The analysis rules out the presence of the proposed planet, as well as of any other planets more massive than 2-20 $\mathrm{M}_\oplus$ $m_{p}$ sin $i$, depending on orbital period. We show that the previously identified long-period RV signal is definitively caused by the magnetic cycle of the star.}


   \maketitle
%-------------------------------------------------------------------

\section{Introduction}

The past three decades have seen an exponential increase in the number of discoveries of exoplanets, reaching more than 6000 known planets and revealing a remarkable diversity of planetary systems beyond our Solar System (Nasa Exoplanet Archive,~\citealt{Christiansen2022}). From hot Jupiters orbiting close to their host stars \citep{MayorQueloz1995} to terrestrial planets in temperate zones (e.g. \citealt{AngladaEscude2016, Gilbert2020}), the characterization of these worlds has become a cornerstone of modern astrophysics. 

Central to this effort is the search for biosignatures in planetary atmospheres that may indicate the presence of life. However, these signs are notoriously difficult to detect and interpret. Current instrumentation and methodology often struggles to obtain the neccesary signal-to-noise ratio to provide strong and univocal detections (e.g.~\citealt{Madhusudhan2025,Taylor2025,Luque2025,Stevenson2025}). Even significant detections would require very careful analysis, as life far from the only source of chemical disequilibrium. 

An alternative approach is the search for technosignatures, i.e. signatures of technology. Such signals, if detected, should be significantly different from those of natural origin, making their interpretation easier. There are several ongoing efforts trying to detect the presence of technosignatures (e.g. SETI,~\citealt{Sagan1975, Slys1985}). However, no clear detection has been reported yet. Some of the more suggestive candidates, such as the \textit{Wow!} signal, have been found to be most likely of astrophysical origin \citep{Mendez2024}.

The lack of clear detections of technosignatures has not stopped more speculative predictions and interpretations of astrophysical data \citep{alien1979,Loeb2022}, and even the discussion on potential threats to Earth \citep{Aliens1986,IndependenceDay1996,AlienResurrection1997,Hibberd2025}. In particular, \citet{alien1979} suggested that a planetary system could exist as the potential origin of a reported transmission coming from $\zeta^2$ Ret, building upon previous similar reports \citep{dickinson1974zeta}. It was later suggested that the origin of this signal might have profound consequences for life on Earth \citep{Aliens1986,AlienVsPredator2004,Prometheus2012}.

$\zeta^2$ Ret is a nearby solar-type star in which no planets have been identified yet. This star has been the target of multiple radial velocity studies in the past, reporting no clear detections \citep{Laliotis2023, Harada2024b}. Despite the lack of a known planetary system, $\zeta^2$ Ret has been identified as a target for the future NASA flagship mission Habitable Worlds Observatory (HWO, \citealt{Mamajek2024}) due to its potential to host undetected Earth-like planets. 

The mistery surrounding $\zeta^2$ Ret, its implications for planetary security, combined with its status as a future target for HWO, makes its one of the most intriguing and promising targets to search for potentially habitable planets. In this work, we study all available data from $\zeta^2$ Ret to assess the potential existence of the planets proposed by \citet{alien1979}, along with any other additional substellar companions.

\section{$\zeta^2$ Ret}

$\zeta^2$ Ret (HD 20807, GJ 138) is a nearby (12.04 pc, \citealt{GaiaEDR3}) G1V star \citep{Keenan1989}, primary of a wide binary pair with $\zeta^1$ Ret. $\zeta^2$ Ret has a mass of 0.98 $\pm$ 0.04 M$_{\odot}$ \citep{Harada2024} and sub-solar metallicity ([FE/H] = -0.215 $\pm$ 0.010, \citealt{Adibekyan2016}). It has an effective temperature of 5846 $\pm$ 59 K and a luminosity \mbox{L$_{*}$ [L$_{\odot}$] = 1.01 $\pm$ 0.06 \citep{Harada2024}}. Using the reported stellar parameters, we estimate the limits of the habitable zone (HZ) of the star to be 0.838 -- 1.774 au, following \citet{Kopparapu2014} for the runaway greenhouse to early-Mars limits. These limits correspond to orbital periods of 295 to 907 days. The star has been proposed to have a magnetic cycle of $\sim$ 7.9 years \citep{Laliotis2023,Harada2024b}. The relevant parameters are shown in Table~\ref{tab:stellar_parameters}.

\begin{table}
   \begin{center}
    \caption{Stellar parameters of $\zeta^2$ Ret.\label{tab:stellar_parameters}}
    \begin{tabular}{lcc}
    \hline\hline
        Parameter                       & Value                    & Ref. \\
        \hline
        RA [J2000]                      & 03:18:12.819 & 1 \\
        DEC [J2000]                     & -- 63:30:22.905 & 1 \\
        $\mu \alpha \cdot \cos\delta$ [mas yr$^{-1}$]& 1331.027 & 1 \\
        $\mu \delta$ [mas yr$^{-1}$]    & 647.725 & 1 \\
        $\Pi$ [$mas$]                   &   83.061 $\pm$ 0.061      & 1 \\ 
        Distance [pc]                   &   12.03925 $\pm$ 0.0087   & 1\\
        $M_G$ [mag]                     &   5.0796 $\pm$ 0.0028    & 1 \\      
        $m_{B}$	 [mag] & 5.814 $\pm$ 0.014  & 2 \\
        $m_{V}$	 [mag] & 5.228 $\pm$ 0.009   & 2 \\
        {[Fe/H]}                        &  --0.215 $\pm$ 0.010        & 3 \\
        T$_{eff}$ [K]       &  5486 $\pm$ 59            & 4 \\
        R$_{*}$ [R$_{\odot}$]            &  0.91 $\pm$ 0.06       & 4 \\
        M$_{*}$ [M$_{\odot}$]             &  0.98 $\pm$ 0.04       & 4 \\
        $\log g$ [cgs] &  4.43 $\pm$ 0.10       & 5 \\
        L$_{*}$ [L$_{\odot}$]    &  1.01 $\pm$ 0.06   & 4 \\
        log$_{10}$R'$_{\rm HK}$ & -- 4.906 $\pm$ 0.039 & 6\\
        P$_{\rm rot ~GP}$ [days] & 20.84 $\pm$ 0.23 & 0 \\
        P$_{\rm cycle}$ [days] &  3007 $\pm$ 32 & 0 \\
        $v$ sin $i$ [km s$^{-1}$] &  1.74 & 7 \\
        Age [Gyr]         &  3.0             & 8 \\
        HZ inner edge [au] & 0.838 $\pm$ 0.024 & 9\\
        HZ outer edge [au] & 1.774 $\pm$ 0.052 & 9\\
        HZ inner period [days] & 295 $\pm$ 15 & 9\\
        HZ outer period [days] & 907 $\pm$ 47 & 9 \\

    \hline
    \end{tabular}
    \end{center}
    \textbf{References:} 0 - This work, 1 -  \citet{GaiaEDR3}, 2 - \citet{Hog2000}, 3 - \citet{Adibekyan2016}, 4 - \citet{Harada2024}, 5 - \citet{Furhmann2015}, 6 - \citet{Lovis2011}, 7 - \citet{Santos2004_Beryllium}, 8 - \citet{Mamajek2008},  9 - Estimated following \citet{Kopparapu2014}. The uncertainties in stellar parameters have been enlarged following \citet{Tayar2022}.
\end{table}

\subsection{Potential presence of planets in the system}

$\zeta^2$ Ret has been a regular target in radial velocity programs aiming at the detection of exoplanets. As shown in section~\ref{obs_data}, almost three decades of RVs are available for this system. However, no planet has been detected so far. Both \citet{Laliotis2023} and \citet{Harada2024b} reported the detection of a long period RV signal which would be consistent with a giant planet in the system. They also reported the detection of a signal of similar period in activity indicators, which prompted them to consider the RV signal to be at least suspicious of being induced by stellar activity. 

\begin{figure}[!ht]
   \centering
  \includegraphics[width=9cm]{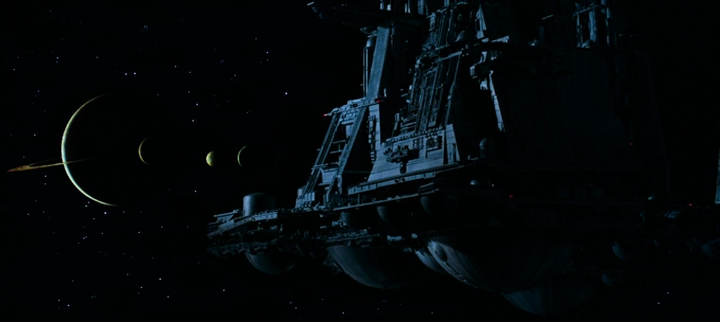}
  \caption{Artist impression of a spaceship arriving Calpamos with its three Earth-like moons in front of it. Taken from \citet{alien1979}}
  \label{calpamos_artist}
\end{figure}

\citet{alien1979} hypothesized the presence of a giant planet orbiting the habitable zone of $\zeta^2$ Ret, hosting a system of at least three Earth-like exomoons (see Fig.~\ref{calpamos_artist}). The planet was later named Calpamos \citep{AliensColonialMarines2013,AliensFireAndStone2014}, with two of its moons named LV-426 \citep{Foster1979} and LV-223 \citep{Prometheus2012}. Not only that, both \citet{alien1979} and \citet{Prometheus2012} discussed the potential detection of technosignatures in these two moons. Calpamos was described as the fourth planet in the $\zeta^2$ Ret, having a mass of $\sim$ 10 $\mathrm{M}_\mathrm{J}$ and orbiting at a distance of $\sim$ 1.08 au from the star, corresponding to an orbital period of $\sim$ 409 days \citep{ColonialMarinesManual1995}. If favorably aligned, the RV signal of this planet should have been already detected \citep{Laliotis2023, Harada2024b}. However, the inclination of the orbital plane of $\zeta^2$ Ret remains unknown, opening the possibility of Calpamos, or a planet of similar characteristics, existing. 

\section{Data} \label{obs_data}

$\zeta^2$ Ret has been extensively observed over the past decades by a plethora of teams using many different high resolution spectrographs. We compiled archival data obtained between 1992 and 2021 from five different instruments. We collected either radial velocity time series, or spectra, from the Coud\'e Echelle Spectrograph and the Long Camera (CES LC, at the 1.4m ESO telescope; \citealt{Enard1982}), its upgraded version CES Very Long Camera (CES VLC, at the 3.6m ESO telescope; \citealt{Kurster1998}), the University College London \'Echelle Spectrograph (UCLES, on the 3.9 Anglo-Australian Telescope; \citealt{Diego1990}), the High Accuracy Radial velocity Planet Searcher (HARPS, on the ESO 3.6m telescope; \citealt{Mayor2003}), and the Planet Finder Spectrometer (PFS, on the 6.5m Magellan Clay telescope; \mbox{\citealt{Crane2006}}).

The RV time series of the CES LC and CES VLC spectrographs were obtained from \citet{Zechmeister2013}, and the UCLES and PFS RV data from \citet{Laliotis2023}. We downloaded the HARPS spectra from the ESO archive and extracted the RVs using the Line-By-Line (\texttt{LBL}\footnote{version 0.65.001, lbl.exoplanets.ca}) code developed by \citet{Artigau2022}, and based on \citet{Dumusque2018}. The \texttt{LBL} algorithm performs an outlier-resistant template matching to each individual line in the spectra. For non-telluric corrected spectra, it produces its own telluric correction. In addition to the velocity, it derives other quantities such as the variations in full width at half maximum of the lines (FWHM), and the differential variations in effective temperature (dTEMP; 
\citealt{Artigau2024_dtemp}).

After nightly binning the data, we obtained a total of 322 RV measurements, distributed as 43 CLES LC RVs, 15 CLES VLC, 99 UCLES, 149 HARPS, and 16 PFS. Both HARPS and PFS underwent interventions that created RV offsets. We split their data as 73 HARPS-03 RVs, 76 HARPS-15 RVs, 13 PFS-0, and 3 PFS-1. We measure a global RV root-mean-square (RMS) of 8.13 m$\cdot$s$^{-1}$, with the RMS of the HARPS data being as low as 2.7 m$\cdot$s$^{-1}$, and a median uncertainty $\sigma$RV of 0.93 m$\cdot$s$^{-1}$. In addition to the RV measurements, we obtained 149 FWHM and dTEMP measurements coming from the HARPS data. The FWHM shows a dispersion of 1.97 m$\cdot$s$^{-1}$ and a median uncertainty of 0.26 m$\cdot$s$^{-1}$. The dTEMP shows a dispersion of 0.129 K, and a median uncertainty of 0.015 K. To avoid numerical issues, we scaled the dTEMP measurements by a factor of 100. Table \ref{tab:data} shows the number of statistics per individual instrument.

\begin{table*}[!ht]
   \begin{center}
       \caption{Spectroscopic data used in this work \label{tab:data}}
       \begin{tabular}[center]{l l l l l l l l}
           \hline
            & CES LC & CES VLC & UCLES & HARPS & PFS & Combined \\
           \hline    
           Extraction & Iodine cell & Iodine cell & Iodine cell & \texttt{LBL} & CCF \\
           N. data & 43 & 15 & 99 & 149 & 16 & 322\\
           Baseline [years] & 5.4 & 5.1 & 17.8 & 17.9 & 7.2 & 28.9\\     
           RV RMS [m$\cdot$s$^{-1}$] & 19 & 10.9 & 4.7 & 2.71 & 2.85 & 8.13\\
           $\sigma$RV [m$\cdot$s$^{-1}$]& 15 & 5.9 & 1.3 & 0.34 & 0.86 & 0.94\\ 
           FWHM RMS [m$\cdot$s$^{-1}$] & &  &  & 1.97 & & 1.97 \\
           $\sigma$FWHM [m$\cdot$s$^{-1}$]& &  &  & 0.26 & & 0.26 \\
           dTEMP [K x100] & & &  &   12.9 & & 12.9 \\
           $\sigma$dTEMP  [K x100]& & &  &  1.5 & & 1.5 \\
           \hline
       \end{tabular}
   \end{center}

\end{table*}

Last, $\zeta^2$ Ret has been observed by the All Sky Automated Survey for SuperNovae (ASAS-SN; \citealt{Kochanek2017}) and the Transiting Exoplanet Survey Satellite (\textit{TESS}) \citep{Ricker2015}. The ASAS-SN data were strongly contaminated by the moon. \textit{TESS} observed $\zeta^2$ Ret in sectors 1, 2, 3, 28, 29, 30, 68, 69, 95, and 96. We found no evidence of transits in the \textit{TESS} data. 

\begin{figure}[!ht]
   \centering
  \includegraphics[width=9cm]{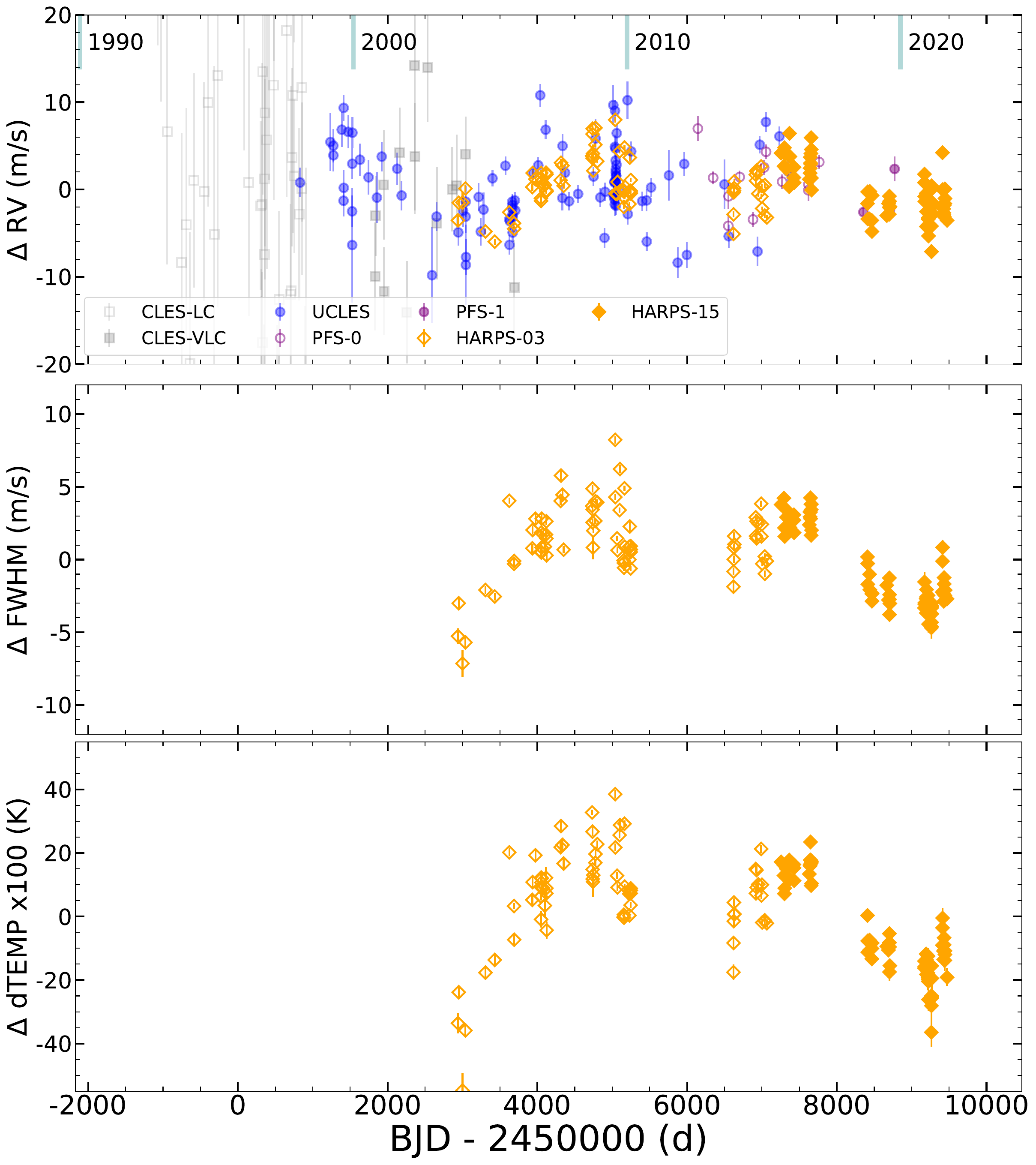}
  \caption{Spectroscopic data used in this work.}
  \label{data}
\end{figure}

\section{Analysis} 

We performed a global analysis of the data following the procedure outlined in \citet{SuarezMascareno2025a, SuarezMascareno2025b}. We modelled all the data at the same time. Every time series includes a zero point per instrument, a model for the cycle, and a model for the stellar rotation. It is the only way to be sure. The cycle model is defined as a combination of two sinusoidal signals with common period and phase for all time series and independent amplitudes. The stellar rotation is modelled as a multi-dimensional Gaussian Process  (GP) \citep{Rajpaul2015}, using the \texttt{S+LEAF} code \citep{Delisle2022} \footnote{\url{https://gitlab.unige.ch/delisle/spleaf}}. Considering the length of the baseline, all time-series include a second order polynomial term to account for potential non-periodic long term variations. In addition, the RV data includes a sine function for each planet in the model. 

We optimised the parameters of the models using Bayesian inference through the nested sampling \citep{Skilling2004, Skilling2006} code \texttt{Dynesty}~\citep{Speagle2020, kosopov2023}. We sampled the parameter space using random slice sampling, which is well suited for the high-dimensional spaces \citep{Handley2015a,Handley2015b} resulting from modelling several time series at once. We used a number of live points of $20 \times N_{free}$ in models with narrow priors in period, and $100 \times N_{free}$ in models with wide priors in frequency, to ensure the discoverability of the narrow frequency posteriors. The significance of any potential detection was assessed through the False inclusion probability (FIP) framework \mbox{\citep{hara2022a}} using the thresholds suggested by \citet{Hara2024}. 

\section{Results and discussion}

\subsection{Activity-only model}\label{presence_act}

We started with an activity-only model, which we define as the null hypothesis. The model includes the cycle, with a prior $\mathcal{LU}[400,5000]$ days. The stellar rotation is modelled with at GP, using a double Stochastic Harmonic Oscillator (SHO) kernel. We used a prior $\mathcal{U}[5,60]$ days for the rotation period, accounting for almost the full range of potential periods of main sequence G-type stars. 

\begin{figure*}[!ht]
   \centering
  \includegraphics[width=16cm]{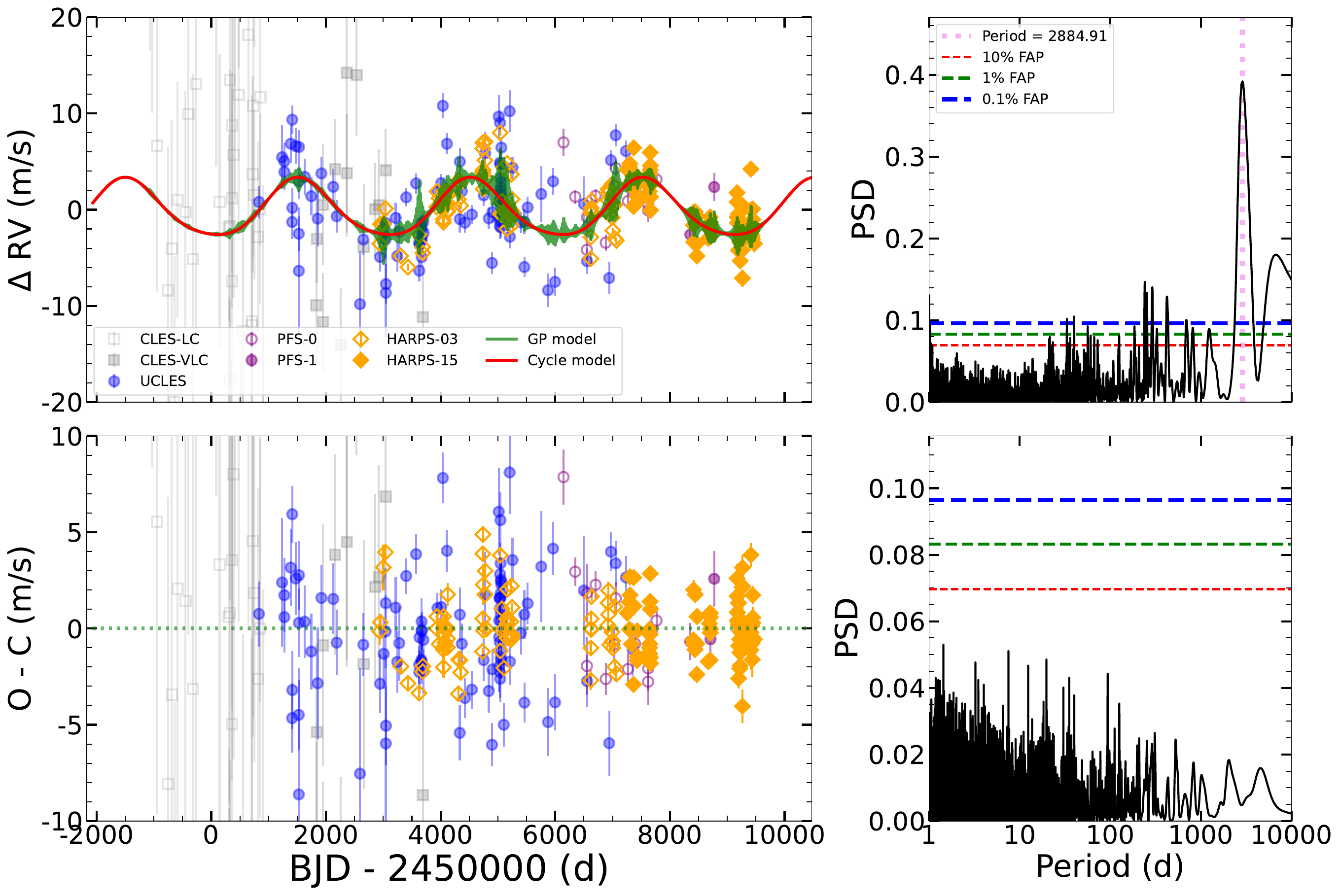}
  \caption{Radial velocity data of $\zeta^2$ Ret with the best model fit (top) and the residuals after the fit (bottom), along with their respective Generalised Lomb Scargle periodograms \citep{Zechmeister2009}.}
  \label{model_vrad}
\end{figure*}

We measured a cycle period of 3007 $\pm$ 32 d, consistent with previous estimates. It has a mild asymmetry and induces an RV-amplitude of 2.93 $\pm$ 0.27 m$\cdot$s$^{-1}$, which would be easy to interpret as an eccentric planetary signal. The values are consistent with those of the RV signals discussed by \citet{Laliotis2023} and \citet{Harada2024b}. Both suspected a stellar origin of the signal, as the period was close to their own measurement of the cycle period. As our model imposes the period and phase to be shared between RV, FWHM, and dTEMP, and no evidence of this long period signal remains in the residuals, we can confirm its origin as a cycle-induced RV signal. The stellar rotation converges to a period of 20.84 $\pm$ 0.23 d, consistent with a star slightly younger than the Sun. All activity signals are well-defined in the three time series.  Fig.~\ref{model_vrad} shows the RV data with the best model fit and fig.~\ref{model_act} shows the same for the activity indicators. The behaviour of the magnetic cycle in FWHM and dTEMP is almost identical, with the shape in RV being almost opposite (see Fig.~\ref{model_cycle}), and very similar to the expected shape of a mildly eccentric Keplerian signal. The residuals after the RV fit showed no evidence of other significant periodic signals (see Fig.~\ref{model_vrad}). 

\subsection{Attempting to detect Calpamos}

To test the presence of the planet Calpamos we included a sinusoid in the model with a prior on the period centered at the value provided by \citet{ColonialMarinesManual1995} with a sigma of 10\% of that value ($\mathcal{LU}[409,41]$ days). We parameterise the planetary amplitude $K$ as ln $K$ with a prior $\mathcal{U}[-5,2.5]$ m$\cdot$s$^{-1}$, allowing for values < 12 m$\cdot$s$^{-1}$, already larger than the RMS of the data. It is obvious from the RMS of the RV dat ($\sim$ 8 m$\cdot$s$^{-1}$) that it is not possible to accommodate the expected RV signal from an edge-on super Jupiter. However, an almost pole-on giant planet could have passed undetected by previous efforts.

We do not obtain a detection of a periodic signal. Instead, we derive an upper limit to potential amplitudes at the test periods of 57 cm$\cdot$s$^{-1}$ (99\% confidence), corresponding to m sin $i$ < 6.4 $\mathrm{M}_\oplus$, far from the suggested mass of 10 $\mathrm{M}_\mathrm{J}$. As inclination plays an important role to the sensitivity, we transform the masses into true masses by assuming a distribution of inclinations cos $i$ $\mathcal{U}[0,1]$, consistent with the known distribution of inclinations of stellar rotation axes. Taking this into account, we estimate an upper limit to the mass of 15.5 $\mathrm{M}_\oplus$. Allowing for some uncertainty in the proposed mass of Calpamos ($\sim$ 20\%), we estimate that an inclination of the orbital plane < 0.13$^{\circ}$ is necessary to explain the results of the model. From the distribution of inclinations, we estimate a probability of 2.6 $\times$ 10$^{-5}$ for the rotation axis to show an inclination < 0.13$^{\circ}$.

\citet{Santos2004_Beryllium} estimated a $v$ sin $i$ of $\sim$ 1.74 km$\cdot$s$^{-1}$ for $\zeta^2$ Ret, with an estimated uncertainty of 0.03 km$\cdot$s$^{-1}$ \citep{Santos2002}. With the rotation period previously determined, and the stellar radius, we estimate the equatorial velocity to be 2.38 $\pm$ 0.12 km$\cdot$s$^{-1}$. This corresponds to an inclination of the rotation axis of 46.9 $\pm$ 3.4$^{\circ}$. If we assumed the orbital plane of the system to be aligned with the rotation axis, then we would compute an upper limit to the mass of 8.9 $\mathrm{M}_\oplus$, and a probability of the inclination being < 0.13$^{\circ}$ of 4 $\times$ 10$^{-7}$.

To further confirm our findings, we performed an injection recovery test in our data. We injected sinusoidal signals using the proposed period and mass of Calpamos (409 d, 10 $\mathrm{M}_\mathrm{J}$, with no uncertainty), an arbitrary time of inferior conjunction, and small inclinations $i$. For inclinations larger than 0.13$^{\circ}$ we get hints of the presence of a signal. For inclinations larger than 0.16$^{\circ}$, corresponding to K$_{\rm RV}$ of 75 cm$\cdot$s$^{-1}$, we can recover the signal with > 3$\sigma$ significance and a median value within 0.2$\sigma$ of the injected value. If a planet like Calpamos exists in the system, we should have found it unless the star was in a strict face-on configuration. 

\begin{figure}[!ht]
   \centering
  \includegraphics[width=9cm]{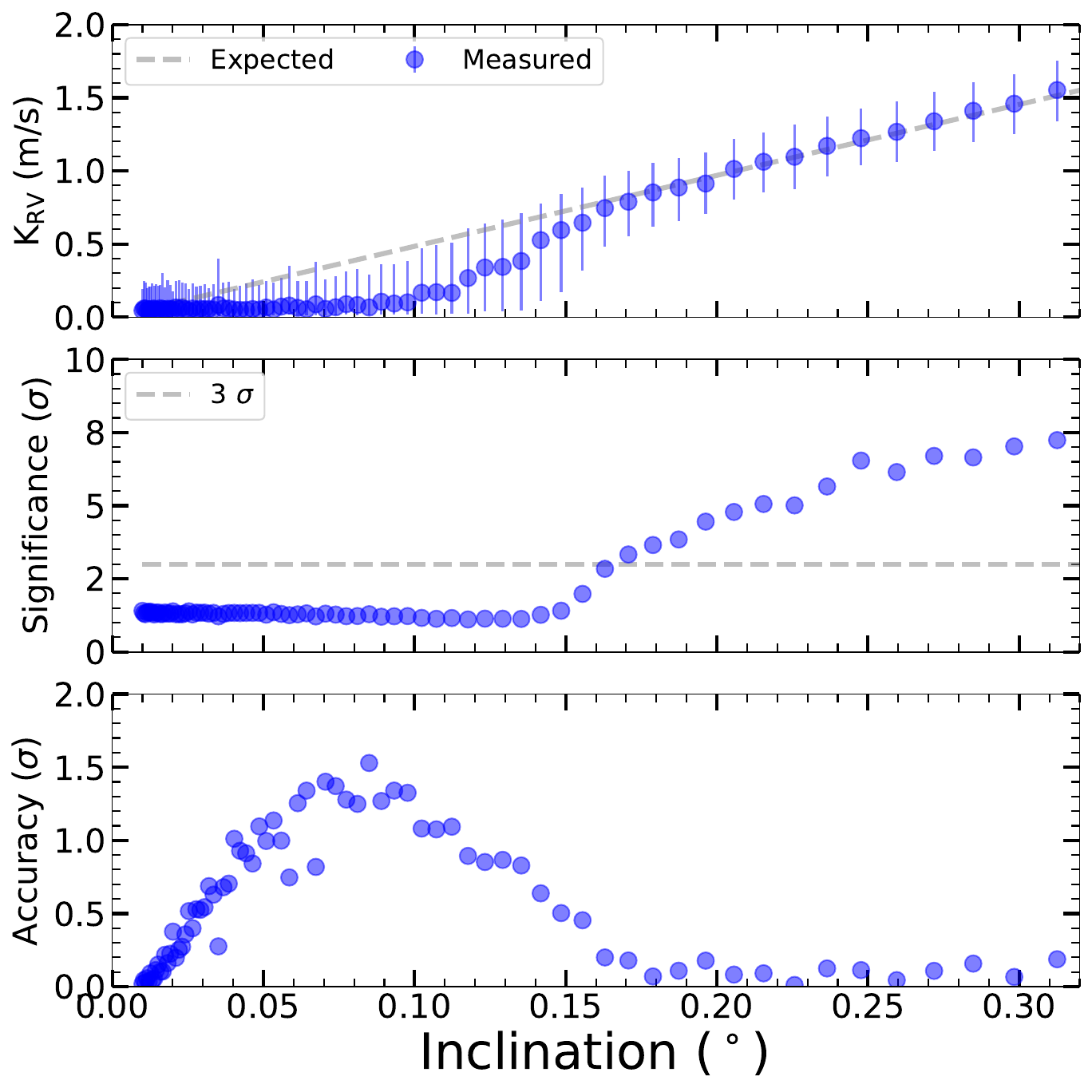}
  \caption{Summary of the injection-recovery test performed to validate upper limits.}
  \label{inj_rec}
\end{figure}

\subsection{Blind search of planets in the system}

To evaluate the potential presence of planets in the system, we ran a blind search using a model with two sinusoidal signals. We used uniform angular frequency priors, corresponding to periods between 2 and 3500 days (1/3 of the observing baseline). We parametrise the time of inferior conjunction as $t_{0} = 2459472 + P_{pl} \cdot (\phi_{pl} - 1)$, with $\phi_{pl}$ parametrised as $\mathcal{U}[0, 1]$. 

\begin{figure}[!ht]
   \centering
  \includegraphics[width=9cm]{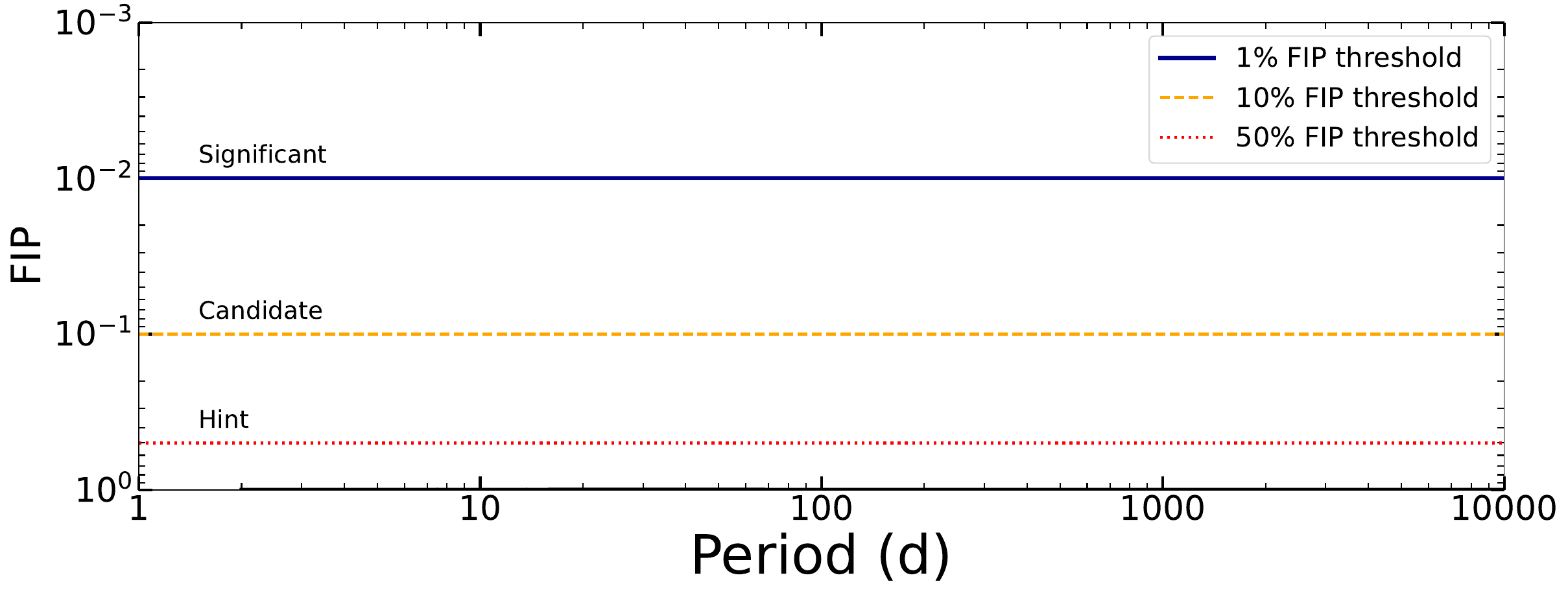}
  \caption{FIP periodogram of the full dataset. No significant power excess is found at any period.}
  \label{fip_periodogram}
\end{figure}

The posterior distribution of the parameters showed no evidence of the presence of signals in the data that could be attributed to planets in the system. We recovered angular frequency posteriors consistent with the full prior, and upper limits for the amplitudes $K$ at 57 cm$\cdot$s$^{-1}$. Figure~\ref{fip_periodogram} shows the FIP periodogram of the full dataset. No power excess rises above 99.5\% FIP.

\subsection{Limits to the presence of planetary companions}\label{signal_stability}

\begin{figure}[!ht]
   \centering
  \includegraphics[width=9cm]{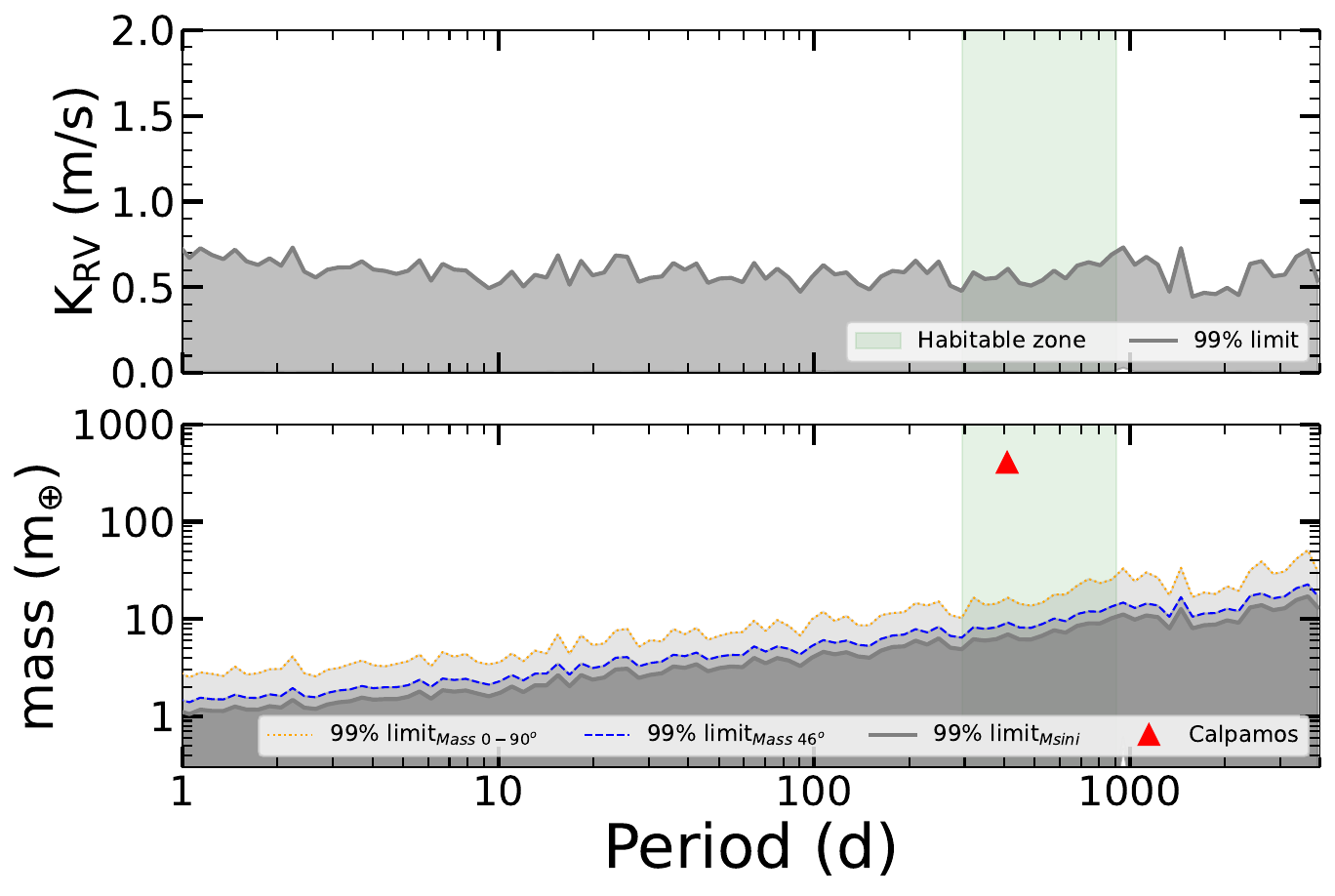}
  \caption{Compatibility limits. The upper panel shows the RV amplitude limit (99\%) as a function of orbital period. The lower panels show the same, but for planetary masses. The green shaded area shows the periods corresponding to the habitable zone. The red triangle shows the 1\% lower limit of the $m_{p}$ sin $i$ of Calpamos.}
  \label{fig_det_lims}
\end{figure}

Finally, to evaluate the periods at which potential planets might exist undetected in the system, we measured the compatibility limits of the data for additional planetary signals in the same way as \citet{SuarezMascareno2025a,SuarezMascareno2025b}. Adopting the result of the activity-only model, we froze most of the parameters of the model (trends, cycle, GP) and left free only the white noise and zero point RV components. We included a sinusoid in the model. We built a grid of 100 bins over periods 1 -- 4000 days with equal width in log$_{10}$ space. We ran this model with a narrow prior for the period within each bin. From the posterior distribution of each try, we computed the 1\% and 99\% limits in RV amplitude, $m_{p}$ sin $i$, and mass, assuming both inclinations of $\sim$ 46$^{\circ}$ and free. Figure~\ref{fig_det_lims} shows the result of this exercise. We measure limits to the amplitudes of potential signals between 0.45 and 0.73 m$\cdot$s$^{-1}$ for all orbital periods. For periods between 1 and 10 days, we found that planets with $m_{p}$ sin $i$ as large as 1.9 $\mathrm{M}_\oplus$ could remain undetected in the system. For periods between 10 and 100 days, planets up to 4.0 $\mathrm{M}_\oplus$. For periods within the habitable zone, planets up to $m_{p}$ sin $i$ 11.1 $\mathrm{M}_\oplus$ could still exist in the system. This would correspond to planets with true masses up to 14.8 $\mathrm{M}_\oplus$ for an inclination of 46$^{\circ}$, or up to 33 $\mathrm{M}_\oplus$ if we do not assume any particular inclination. Overall, we find it unlikely that $\zeta^2$ Ret could host any planet with an $m_{p}$ sin $i$ larger than 22.7 $\mathrm{M}_\oplus$, or a true mass larger than 51 $\mathrm{M}_\oplus$, at periods shorter than 4000 days. The presence of giant planets (e.g. Jupiter mass) is only possible at face-on configurations. We identified a power excess, at a period of $\sim$ 950 days, where the 1\% limit in RV amplitude and $m_{p}$ sin $i$ are not consistent with zero. More data would be needed to confirm whether this power excess is due to a potential planetary signals. Considering the period and potential amplitude, projects undertaking this task would need to make a very significant and risky effort. We cannot lie about their chances, but they have our sympathies.

\begin{table}
   \begin{center}
    \caption{Limits to the presence of planetary companions.  \label{tab:limits}}
    \begin{tabular}{l l l l l}
    \hline
       Period range & K$_{RV}$           & $m_{p}$ sin $i$     & mass$_{~46^{\circ}}$ & mass\\
       (d)          & (m/s)               & ($\mathrm{M}_\oplus$)  & ($\mathrm{M}_\oplus$)  & ($\mathrm{M}_\oplus$) \\
    \hline
     1 -- 10 & 0.73 & 1.9 & 2.4 & 4.6 \\
     10 -- 100 &  0.68 & 4.0 & 5.3 & 9.9\\
     100 -- 1000 & 0.73 & 11.1 & 14.8 & 33.1\\
     1000 -- 4000 & 0.73 & 17.1 & 22.7 & 51.0 \\
     295 (HZ$_{inner}$) & 0.59 & 6.2 & 8.2 & 16.7\\
     907 (HZ$_{outer}$) & 0.73 & 11.1 & 14.8 & 33.1\\
     \hline
    \end{tabular}
    \end{center}
    Note: The values displayed are the largest within the period range.
\end{table}

We explored the possibility of complementing these limits using Gaia astrometry. $\zeta^2$ Ret has a renormalized unit weight error (RUWE) of 1.202 \citep{GaiaEDR3}, suggesting a solution consistent with a single star and limiting the astrometric study of the system. We estimated that a planet as large as the hypothesized Calpamos would produce an astrometric effect of 0.87 mas, significantly lower than the precision of along-scan astrometry of the Gaia mission for a star of its brightness \citep{Lindegren2018}.

\section{Conclusions}

We analysed 28.8 years of radial velocities of $\zeta^2$ Ret with the aim of investigating the existence of the giant planet Calpamos, suggested by \citet{alien1979}. We found the presence of the planet Calpamos to be incompatible with the data, except for a strict face-on configuration ($i$ < 0.13$^{\circ}$). A blind search of planets in the system did not reveal the presence of any planet at orbital periods shorter than 4000 days. 

Our analysis showed the long-period RV signal described in \citet{Laliotis2023} and \citet{Harada2024b} to be induced by the magnetic cycle of the star by showing that the signal in RV shares the same period and phase with the equivalent signal in activity indicators. The cycle has a period of 3007 $\pm$ 32 d and induces an RV amplitude of 2.93 $\pm$ 0.27 m$\cdot$s$^{-1}$. In addition, we determined the rotation period to be 20.84 $\pm$ 0.23 d. 

The RV data of $\zeta^2$ Ret is incompatible with the presence of planets with $m_{p}$ sin $i$ > 17.1 $\mathrm{M}_\oplus$ at any orbital period (< 4000 d), with the limits around the habitable zone ranging from 6.2 $\mathrm{M}_\oplus$ to 11.1 $\mathrm{M}_\oplus$. These limits correspond to a limit on the RV amplitude between 0.73 m$\cdot$s$^{-1}$ over the full period baseline. Using the distribution of inclinations of the rotation axes of stars, we estimate it unlikely that planets with a mass < 51 $\mathrm{M}_\oplus$ could exist in the system (99\% confidence). Nevertheless, we cannot rule the presence of larger planets in strict face-on configurations.

\begin{acknowledgements}

A.S.M. acknowledges financial support from the Spanish Ministry of Science and Innovation (MICINN) project PID2020-117493GB-I00 and from the Government of the Canary Islands project ProID2020010129. 

\newline
This work is based on data obtained via the  HARPS public database at the European Southern Observatory (ESO). We are grateful to all the observers of the following ESO projects, whose data we are using: 072.C-0488, 072.C-0513, 073.C-074, 074.C-0012, 076.C-0878, 077.C-0530, 078.C-0833, 079.C-0681, 183.C-0972, 192.C-0852, 196.C-1006, 0102.C-0584, 0103.C-0206, 106.215E, 105.20AK. We are grateful to the crews at the ESO observatory of La Silla, the Las Campanas Observatory, and the Australian Astronomical Observatory. This research has made extensive use of the SIMBAD database, operated at CDS, Strasbourg, France, and NASA's Astrophysics Data System. This research has made use of the NASA Exoplanet Archive, which is operated by the California Institute of Technology, under contract with the National Aeronautics and Space Administration under the Exoplanet Exploration Program.

\newline
The manuscript was written using \texttt{VS Code}. 
Main analysis performed in \texttt{Python3} \citep{Python3} running on \texttt{Ubuntu} \citep{Ubuntu} systems and \texttt{MS. Windows} running the \texttt{Windows subsystem for Linux (WLS)}.
Extensive usage of \texttt{Numpy} \citep{Numpy}.
Extensive usage of \texttt{Scipy} \citep{Scipy}.
All figures built with \texttt{Matplotlib} \citep{Matplotlib}.
The bulk of the analysis was performed on desktop PC with an AMD Ryzen$^{\rm TM}$ 9 9950X (16 cores, 32 threads, 3.5--4.7 GHz), provided by ASM.

\newline
This article was inspired by rewatching Alien and thinking \textit{wait, I know that star, it has HARPS data!}.

\end{acknowledgements}

\bibliography{biblio}

\begin{appendix}

\clearpage

\onecolumn

\section{Additional figures}

\begin{figure*}[!ht]
   \centering
  \includegraphics[width=18cm]{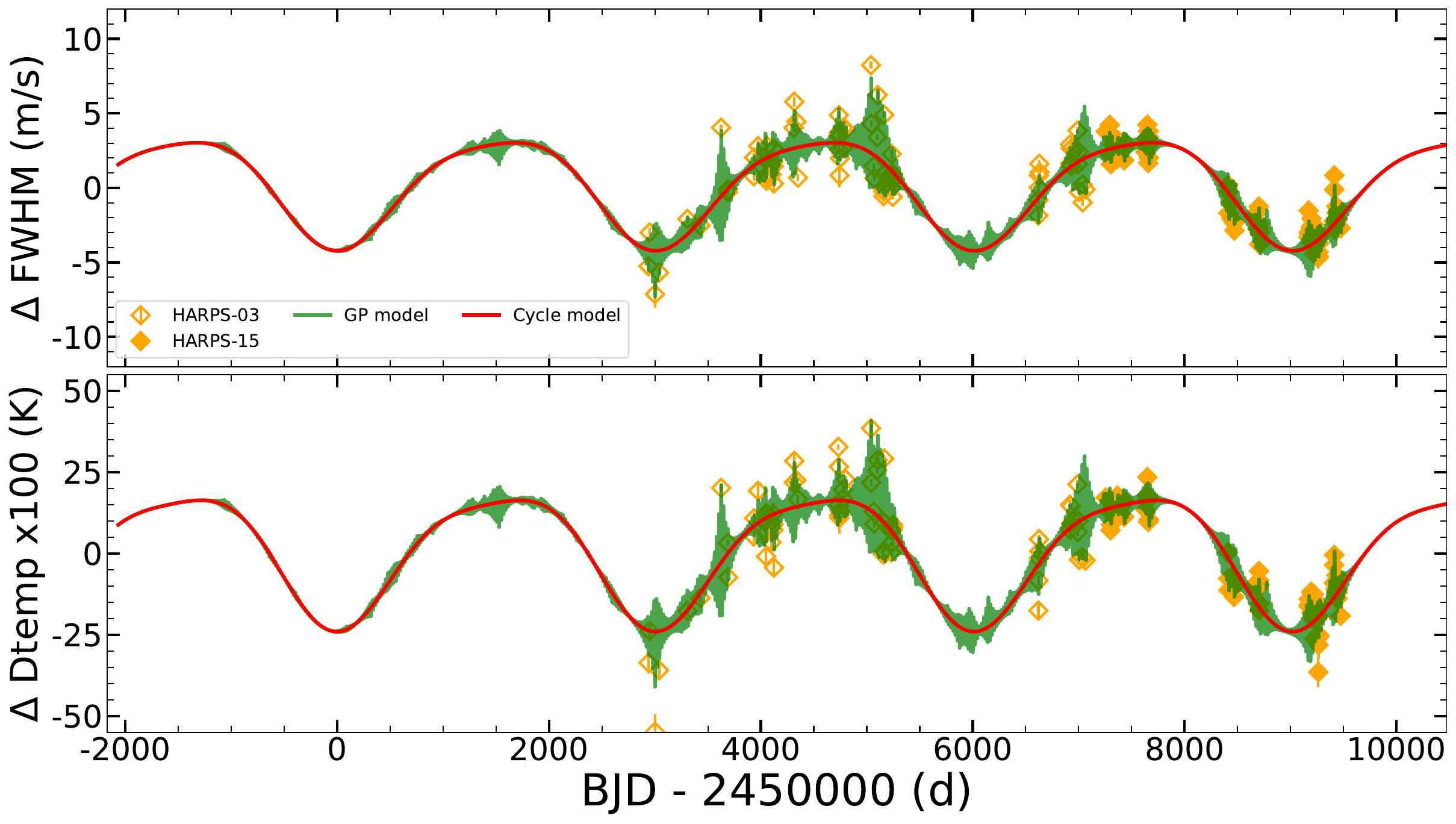}
  \caption{Times series data of FWHM (top) and dTEMP (bottom), along with the best model fit.}
  \label{model_act}
\end{figure*}

\begin{figure*}[!ht]
   \centering
  \includegraphics[width=18cm]{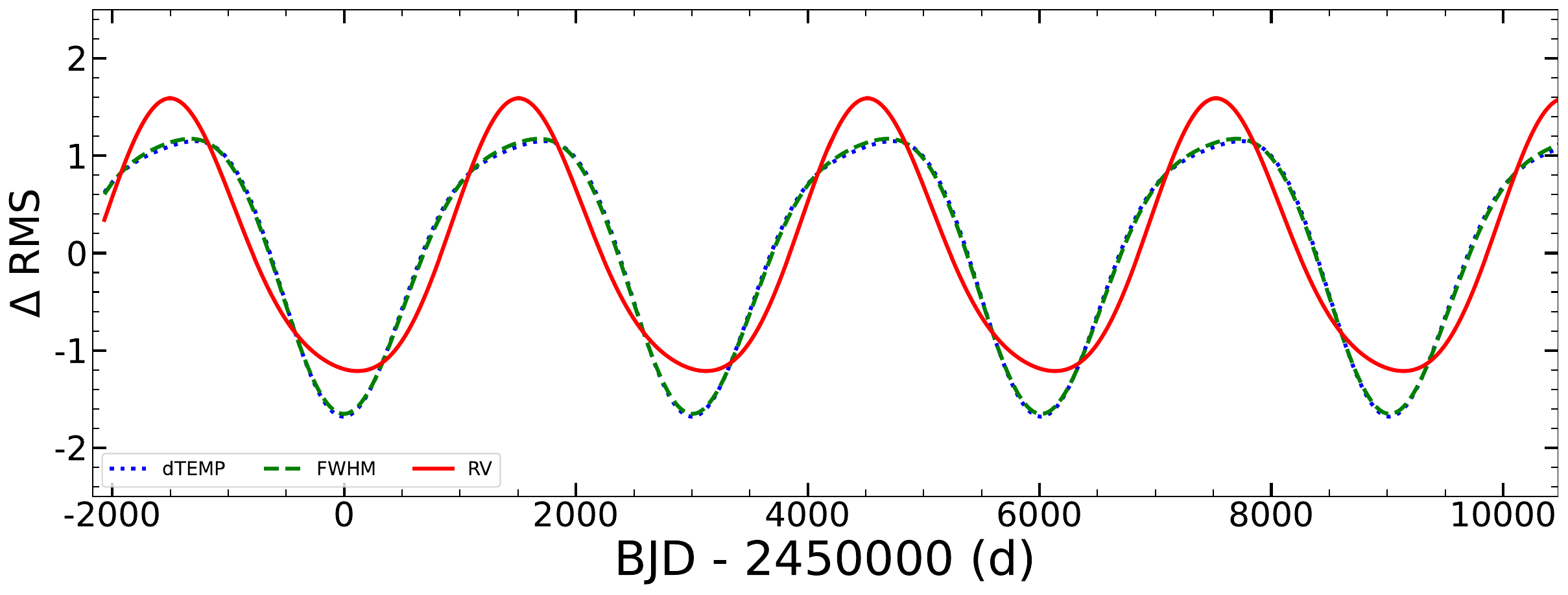}
  \caption{Comparison of the cycle models in RV, FWHM, and dTEMP. All amplitudes have been normalized to the RMS of their respective models.}
  \label{model_cycle}
\end{figure*}

\end{appendix}

\label{lastpage}

\end{document}